\documentclass[conference, 9pt]{IEEEtran}
\IEEEoverridecommandlockouts

\usepackage{setspace}
\usepackage{float}
\usepackage{dblfloatfix}
\usepackage{cite}
\usepackage{amsmath,amssymb,amsfonts}
\usepackage{algorithmic}
\usepackage{graphicx}
\usepackage{textcomp}
\usepackage{caption}
\usepackage{subcaption}
\usepackage{xcolor}
\usepackage[a4paper, total={184mm,239mm}]{geometry}

\usepackage{enumitem}
\usepackage[font=footnotesize]{caption}
\usepackage{tikz}

\usepackage{fancyhdr}
\newcommand*\circled[1]{\tikz[baseline=(char.base)]{
            \node[shape=circle,fill,inner sep=1pt] (char) {\textcolor{white}{#1}};}}

\makeatletter
\renewcommand\footnoterule{
  \kern-3\p@
  \hrule\@width 0.5\columnwidth
  \kern2.6\p@}
  \makeatother

\newcommand{\floor}[1]{\lfloor #1 \rfloor}

\def\BibTeX{{\rm B\kern-.05em{\sc i\kern-.025em b}\kern-.08em
    T\kern-.1667em\lower.7ex\hbox{E}\kern-.125emX}}
    
\renewcommand{\IEEEbibitemsep}{0.3pt}
\makeatletter
\IEEEtriggercmd{\reset@font\normalfont\fontsize{7.2pt}{7.5pt}\selectfont}
\makeatother
\IEEEtriggeratref{1}

\begin{document}

\title{DNN-Life: An Energy-Efficient Aging Mitigation Framework for Improving the Lifetime of On-Chip Weight Memories in Deep Neural Network Hardware Architectures \vspace{-10pt}}

\author{\IEEEauthorblockN{Muhammad Abdullah Hanif$^1$, Muhammad Shafique$^2$}
\IEEEauthorblockA{\textit{$^1$Faculty of Informatics, Technische Universit{\"a}t Wien (TU Wien), Vienna, Austria} \\
\textit{$^2$Division of Engineering, New York University
Abu Dhabi (NYUAD), Abu Dhabi, United Arab Emirates}\\
muhammad.hanif@tuwien.ac.at, muhammad.shafique@nyu.edu}}

\maketitle
\thispagestyle{fancy}
\begin{abstract}
Negative Biased Temperature Instability (NBTI)-induced aging is one of the critical reliability threats in nano-scale devices. This paper makes the first attempt to study the NBTI aging in the on-chip weight memories of deep neural network (DNN) hardware accelerators, subjected to complex DNN workloads. We propose DNN-Life, a specialized aging analysis and mitigation framework for DNNs, which jointly exploits hardware- and software-level knowledge to improve the lifetime of a DNN weight memory with reduced energy overhead. At the software-level, we analyze the effects of different DNN quantization methods on the distribution of the bits of weight values. Based on the insights gained from this analysis, we propose a micro-architecture that employs low-cost memory-write (and read) transducers to achieve an optimal duty-cycle at run time in the weight memory cells, thereby balancing their aging. As a result, our DNN-Life framework enables efficient aging mitigation of weight memory of the given DNN hardware at minimal energy overhead during the inference process.  
\vspace{-5pt}
\end{abstract}

\section{Introduction}
\label{Sec1:Introduction}

\begin{figure}[!t]
\centering
\includegraphics[width=1\linewidth]{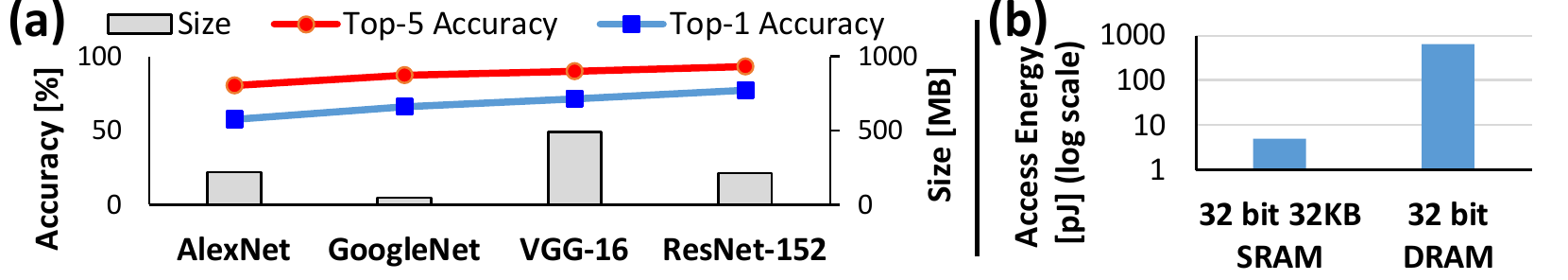}
\caption{(a) Accuracy and size comparison of few of the state-of-the-art DNNs (b) Access energy comparison of SRAM with DRAM (data source:~\cite{Sze_Survey}).}
\label{fig:1a}
\end{figure}

DNN accelerators have already become an essential part of various machine learning systems~\cite{Sze_Survey}\cite{capra2020hardware}. 
DNNs usually require a large number of parameters to offer high accuracy, which comes at the cost of high memory requirements; see Fig.~\ref{fig:1a}a. 
Dedicated memory hierarchies are designed to tradeoff between the low-cost storage offered by the off-chip DRAMs and the energy-/performance-efficient access offered by the on-chip SRAMs~\cite{Sze_Survey}; see Fig.~\ref{fig:1a}b for access energy. 
This has led to an increasing trend towards the use of larger on-chip memory in the state-of-the-art DNN accelerators~\cite{chen2014dadiannao}\cite{jouppi2017datacenter}, with the recent wafer-scale chips having up to 18 GBs of on-chip memory~\cite{HOTCHIP}. 
However, due to continuous technology scaling, the on-chip SRAM-based memories are becoming increasingly vulnerable to different reliability threats, for example, soft errors and aging~\cite{henkel2013reliable}\cite{shafique2020robust}\cite{henkel2013thermal}. 
Studies have shown that even a single fault in weights of critical neurons can result in significant degradation of application-level accuracy~\cite{8474192}. 
State-of-the-art works have focused on analyzing and mitigating the effects of faults in DNN accelerators w.r.t. DNN accuracy~\cite{kim2018matic}. 
\textit{However, to the best of our knowledge, no prior works have analyzed and optimized the aging of the on-chip weight memories of DNN accelerators, especially when considering diverse dataflows of different DNNs and the impact of different types of quantizations on the weight distributions.} 

\textbf{Aging due to NBTI:} In PMOS transistors when a negative gate-to-source voltage is applied, it can break-down the Si-H bond at the oxide-interface, thereby causing a gradual increase in the threshold voltage ($V_{th}$) over the device lifetime, which results in poor drive current and a reduction in the noise margin~\cite{kang2008nbti}\footnote{A similar phenomenon called PBTI happens in NMOS transistors, though NBTI has been considered relatively more serious compared to PBTI~\cite{henkel2013reliable}.}. 
To overcome this $V_{th}$ shift, the operating frequency of the device has to be reduced by more than 20\% over its entire lifetime~\cite{gnad2015hayat}. 
However, due to strict performance and energy constraints (specifically for embedded applications), the $V_{th}$ shift cannot be addressed just by design-time delay margins or adaptive operating frequency adjustments~\cite{shin2008proactive}, as this leads to a significant loss in the system's performance and energy efficiency. 
Therefore, in traditional computing systems, alternate opportunities have to be exploited to overcome this challenge~\cite{gnad2015hayat}. 
One such opportunity lies in the fact that the NBTI aging phenomenon is partially reversed by removing the stress. 

\begin{figure}[t]
\centering
\includegraphics[width=0.95\linewidth]{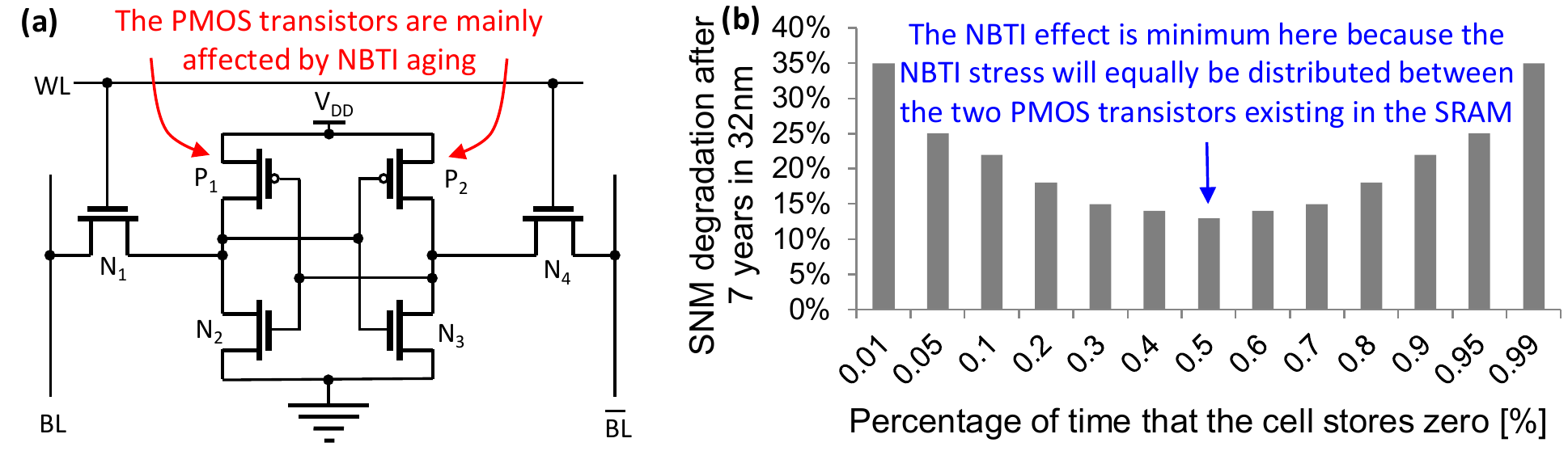}
\caption{(a) A 6T-SRAM Cell; and (b) its SNM degradation after 7 years~\cite{kothawade2011analysis}}
\label{fig:1}
\end{figure}

\textbf{NBTI Aging of On-chip Memories:} 
On-chip memories are typically built using 6T-SRAM cells to achieve high area and power efficiency. A 6T-cell is composed of two inverters coupled with two access transistors (see Fig.~\ref{fig:1}a). The inverters store complementary values to store a single bit. 
Each inverter has a PMOS transistor and an NMOS transistor. Depending on whether the cell is storing `0' or `1', one of the PMOS transistors is always under stress, when the transistor is on. 
As aging of a cell is defined by its most-aged transistor, the lowest aging is achieved when both the PMOS transistors receive on-average the same amount of stress over the entire lifetime of the device, i.e., the percentage of the entire lifetime for which the cell stores a `1' (\textit{duty-cycle}) is 50\%, as shown in Fig.~\ref{fig:1}b. 
Note that NBTI aging strongly depends on average long-term stress and weakly on short-term statistics~\cite{abella2007penelope}. \textit{Therefore, the key challenge in aging mitigation of on-chip memories is to balance their duty-cycle over the entire lifetime without affecting system-level performance.} 

\textbf{State-of-the-art techniques and their limitations:}
Various techniques have been proposed at circuit-level and at architecture-level. At circuit-level, the structure of SRAM-cells is modified to reduce the aging rate~\cite{ricketts2010investigating}\cite{shin2008proactive}. 
For example, Ricketts et al.~\cite{ricketts2010investigating} proposed an asymmetric SRAM structure for workloads having biased bit distribution, but due to their high data dependence, they are applicable only in specific scenarios. Recovery boosting through dedicated recovery accelerating circuit is another method for enhancing the lifetime of the SRAM cells~\cite{siddiqua2011enhancing}, but it increases power/energy consumption due to additional transistors per cell, and therefore cannot be used in energy-constrained large-sized memories~\cite{6105303}. At architecture-level, periodic inversion of data is used to reduce the aging rate of on-chip caches~\cite{jin2012aging}. However, it cannot guarantee optimal duty-cycle, specifically in cases where the same data is periodically reused, e.g., in DNN-based systems where the same set of parameters are reused for processing each input sample. Calimera et al. in~\cite{calimera2011partitioned} improved recovery of unutilized portions of memory, but at high area \& energy cost of expensive online monitoring. The technique also suffers from serious performance degradation in dynamic workload scenarios. Another set of techniques uses bit rotations to cater NBTI aging in registers~\cite{kothawade2011analysis}, but they work only in cases where the overall distribution of bits is relatively balanced. Moreover, they use barrel shifters that incur high area and power overheads. The work in~\cite{Shafique:2015:EEA:2744769.2744834} proposed a configurable micro-architecture for reducing aging rate of video memories, but only works for streaming video applications. 

\textit{In summary,} the state-of-the-art techniques either incur high overheads in terms of area and power/energy or rely on certain specific workloads, but cannot be employed in DNN accelerators due to the unique properties of DNN hardware and workloads, as we will illustrate later in this paper.

\textbf{Additional Challenges from the Deep Learning Perspective:} The dataflow (i.e., computation scheduling) for a given DNN on a specific hardware is defined as per the DNN architecture and the hardware implementation to achieve maximum energy-/performance-efficiency. Altering the dataflow to balance the duty-cycle in on-chip SRAM cells can result in significant degradation of system-level efficiency. \textit{Therefore, an aging mitigation technique that does not require any alteration to the dataflow or the mapping of the data in on-chip SRAM is desired.} 

\begin{figure}[h]
\centering
\includegraphics[width=1\linewidth]{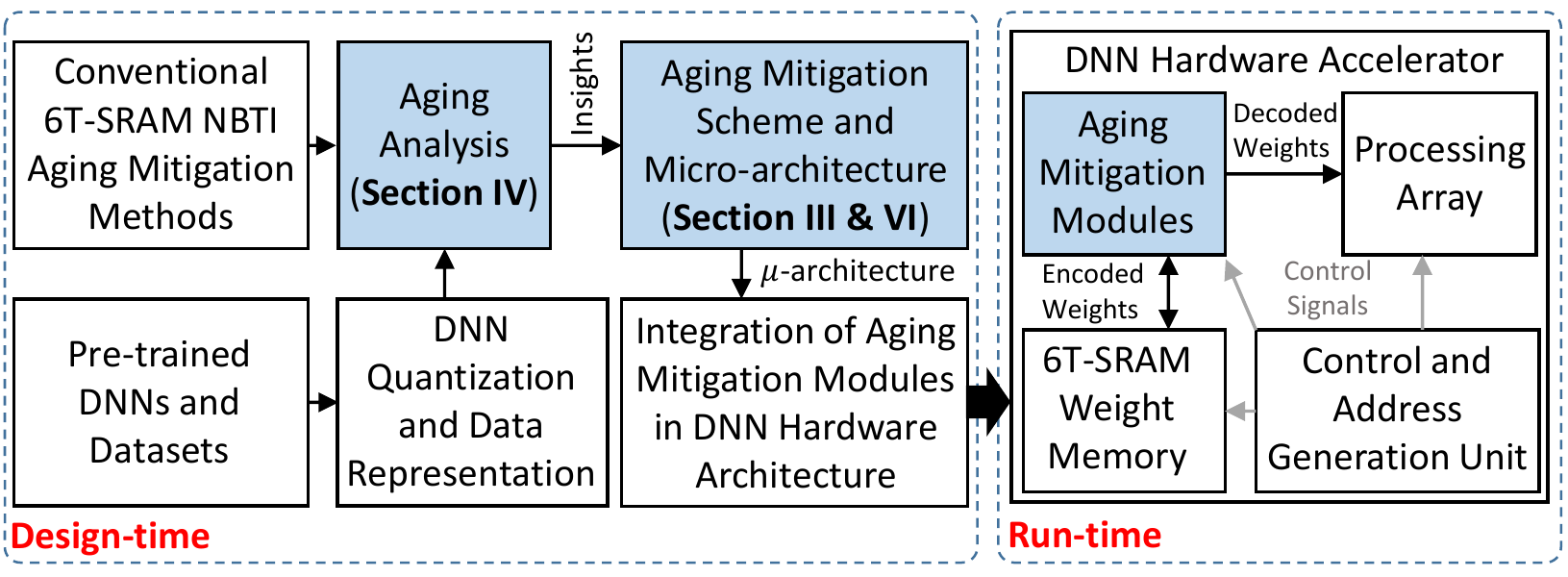}
\caption{Overview of the design-time and run-time steps involved in our DNN-Life framework. Our novel contributions are highlighted in colored boxes.\vspace{-5pt}}
\label{fig:novel_contribution}
\end{figure}

\textbf{Our Novel Contributions:}
Towards this, we propose DNN-Life, an aging analysis and mitigation framework for on-chip memories of DNN hardware (see Fig.~\ref{fig:novel_contribution}). Our framework employs two key features: 
\begin{enumerate}[leftmargin=*]
    \item \textit{Aging Analysis [Section~\ref{sec:analysis}]:} We analyze the impact of using different data representation formats and quantization methods for weights of a DNN on the probability distribution of weight-bits, as this can provide useful insights for designing an effective and low-overhead aging mitigation technique. 
    \item \textit{Aging Mitigation [Section~\ref{sec:novel_micro_architecture}]:} We propose a scheme and supporting micro-architecture for mitigating the NBTI-aging of 6T-SRAM-based on-chip weight memory of DNN accelerators with minimal energy overhead. Noteworthy, our scheme does not require any alteration to the dataflow of DNN inference or on-chip data mapping, and thereby maintains the energy and performance benefits of the system. The micro-architectural extensions for aging mitigation are integrated in the DNN accelerator before and after the on-chip weight memory in the form of aging-aware write and read transducers, as shown in Fig.~\ref{fig:5a}. \vspace{-5pt}
\end{enumerate}

\section{Overview of Our DNN-Life Framework}
\label{sec:Overview_section}

\begin{figure*}[t]
\centering
\includegraphics[width=1\linewidth]{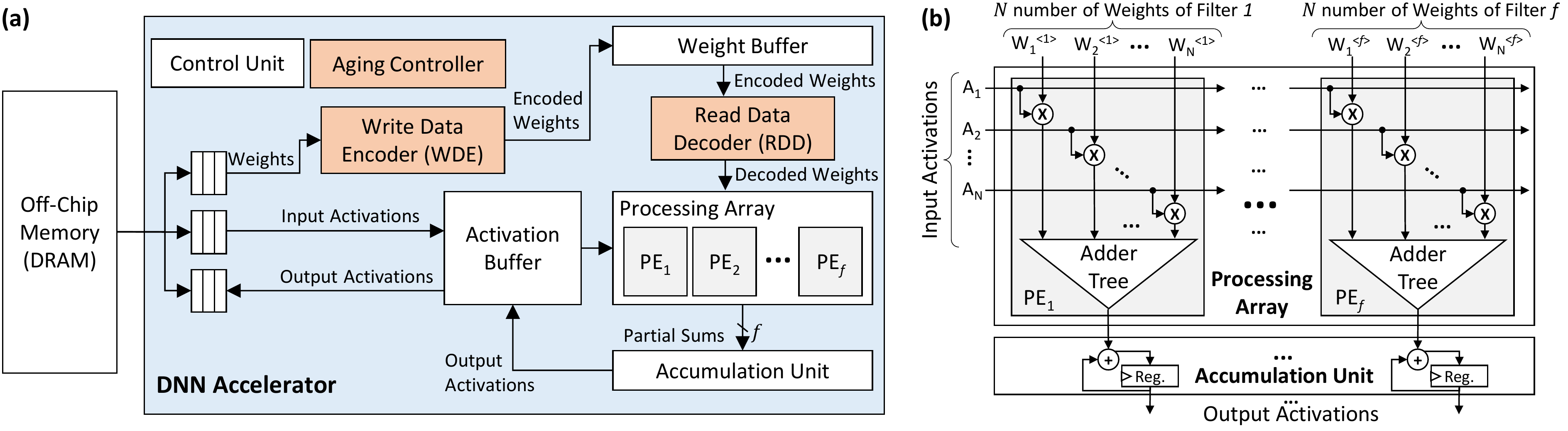}
\caption{(a) Architecture of the baseline DNN accelerator.  The highlighted boxes, i.e., Write Data Encoder (WDE), Read Data Decoder (RDD) and Aging Controller, are the proposed modules for mitigating NBTI aging of weight memory. (b) A detailed view of the processing array and the accumulation unit. }
\label{fig:5a}\vspace{-10pt}
\end{figure*}

In this work, we propose DNN-Life, a novel aging analysis and mitigation framework for weight memories of DNN hardware accelerators. It employs a low-cost data encoding scheme that accounts for diverse DNN workloads to adapt over time to balance the duty-cycle in each on-chip weight memory cell to alleviate the NBTI-aging effects. Towards this, the two key features of our framework are: 
\begin{enumerate}[leftmargin=*]
    \item \textbf{Analysis:} We analyze the probability distribution of weight-bits of different pre-trained DNNs to find key insights that help in developing a low-cost aging-mitigation scheme. To consider the variations in the distribution across number representation formats and the methods used to transform the weights to those formats, we consider different number representation formats and different commonly used conversion methods. The detailed analysis and insights are presented in Section~\ref{sec:analysis}.
    \item \textbf{Architecture:} Based on the gathered insights, we design a data encoding module and an aging controller. The encoder is responsible for encoding the weights before writing the values to the weight memory, and the aging controller is responsible for generating encoding information required to encode the data such that the duty-cycle is balanced. The encoding information is then stored to be used by the corresponding decoder module. The data encoder is deployed inside the DNN hardware accelerator right before the weight memory, and the corresponding decoder is installed after the memory, to decode the weights before passing them for computations. The integration of the encoder and the decoder modules in a DNN accelerator is illustrated in Fig.~\ref{fig:5a}a. The details of the micro-architecture are presented in Section~\ref{sec:novel_micro_architecture}.
\end{enumerate}

\subsection{DNN Hardware Architecture}
\label{sec:baseline_hardware_architecture}
Our DNN hardware architecture is based on well-established DNN accelerator models, such as~\cite{delmas2018bit} for dense DNNs.
Our accelerator is composed of an \textit{Activation Buffer}, a \textit{Weight Buffer}, a \textit{Processing Array}, and an \textit{Accumulation Unit}; see Fig.~\ref{fig:5a}a. 
Our proposed weight-memory aging mitigation modules integrated in the architecture are also shown in the figure (see details in Section~\ref{sec:novel_micro_architecture}). 
The activation and weight buffers provide intermediate storage for the activations and weights, respectively, to reduce the costly off-chip memory accesses. 
The buffers provide data to the processing array for performing the computations. 
For this work, we assume a memory hierarchy similar to Bit-Tactical~\cite{delmas2018bit}, DaDianNao~\cite{chen2014dadiannao} and TPU~\cite{jouppi2017datacenter}, according to which: 1) the activation buffer is large enough to store the activations of a single layer of a DNN; 2) the activation memory can provide $N$ number of activation values to the processing array at a time; and 3) the weight memory can provide $f \times N$ weights to the processing array simultaneously. The processing array (see Fig.~\ref{fig:5a}b) is composed of $f$ number of \textit{Processing Elements} (PEs) that share the activations, and therefore can perform $N$ number of multiplications for $f$ different filters at the same time. Each PE has an adder tree to compute the sum of the multiplications. The computed sum is passed to the accumulation unit where it is added with the corresponding partial sums to generate the output activation value. Note, as the filters can be significantly large, the computation of each output activation can take several cycles, depending on the filter size. 

\begin{figure}[t]
\centering
\includegraphics[width=1\linewidth]{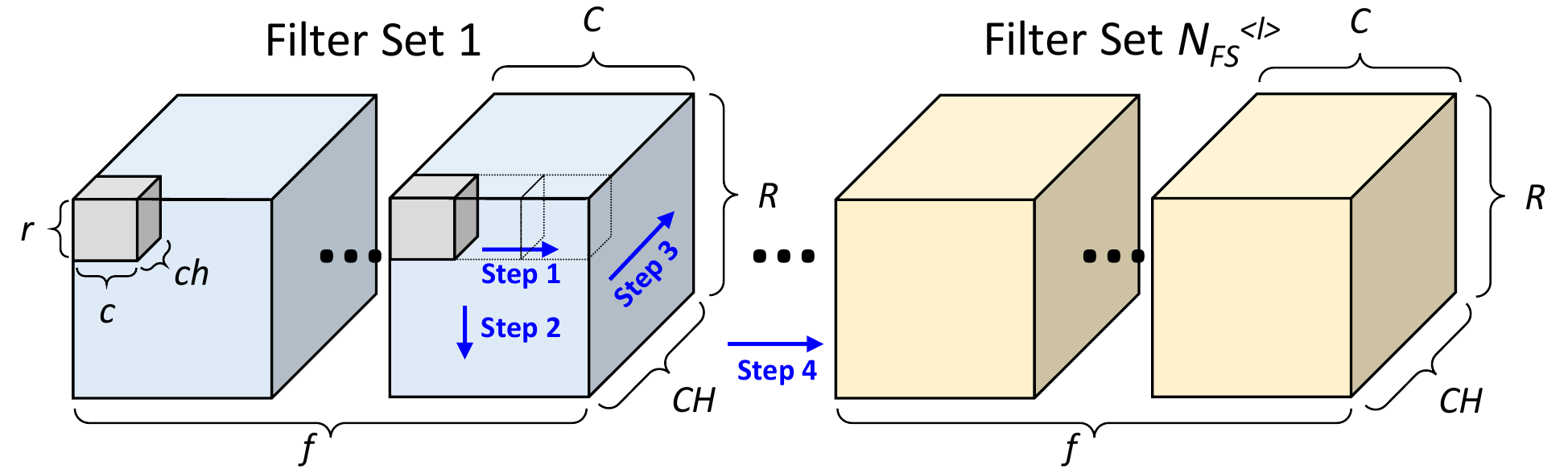}
\caption{Division of filters of a CONV layer of a DNN into smaller blocks that can be accommodated in the on-chip weight memory. Different colors correspond to different sets of filters/blocks. The gray colored boxes define one block of $r\times c\times ch\times f$ size. The \textit{steps} show the sequence in which the blocks are moved to the on-chip fabric for scheduling their computations.}
\label{fig:Dataflow}
\end{figure}

\subsection{Dataflow in the DNN Accelerator}
To perform the computations of a DNN layer using the above accelerator, the weights have to be partitioned into blocks that can be accommodated in the on-chip memory. The goal of partitioning is to maximize the use of available PEs. 
The input/output feature maps and the filters/neurons all are divided into so-called \textit{tiles}, depending on the available on-chip storage for the corresponding data type. 
Works like SmartShuttle~\cite{8342033} provide methods to find an optimal tiling configuration and computation scheduling policy for a layer of a DNN for a given memory hierarchy. 

Fig.~\ref{fig:Dataflow} illustrates the policy that we employ for partitioning the filters of a CONV layer. Note, we support the well-established tiling technique so that we can demonstrate that our technique can benefit a wide-range of existing DNN hardware accelerators.  
The figure also illustrates the sequence in which the blocks are moved to the on-chip weight memory and the corresponding computations are scheduled. 
The filters are first divided into sets, where each set contains \textit{f} number of filters. 
Note, \textit{f} is mainly defined based on the number of filters that the hardware accelerator can process in parallel. 
Afterwards, a chunk of data (grey boxes in Fig.~\ref{fig:Dataflow}) from a set is selected to be moved to the on-chip memory. 
The selected chunk contains a block of data of size $r \times c \times ch$ from the same location of each filter in the set. 
The sequence in which the grey boxes are traversed in the filters defines rest of the dataflow. 
The used sequence is shown as steps in Fig.~\ref{fig:Dataflow}.

\section{Analysis of the Distribution of Weight-Bits for Different DNNs \& their Impact on Duty-Cycle}
\label{sec:analysis}

Before presenting the design of the proposed aging mitigation modules in Section~\ref{sec:novel_micro_architecture}, here we first present an analysis which highlights the rationale behind the proposed design. 

\subsection{Analyzing the Distribution of Weight-Bits}
\begin{figure}[ht]
\centering
\includegraphics[width=1\linewidth]{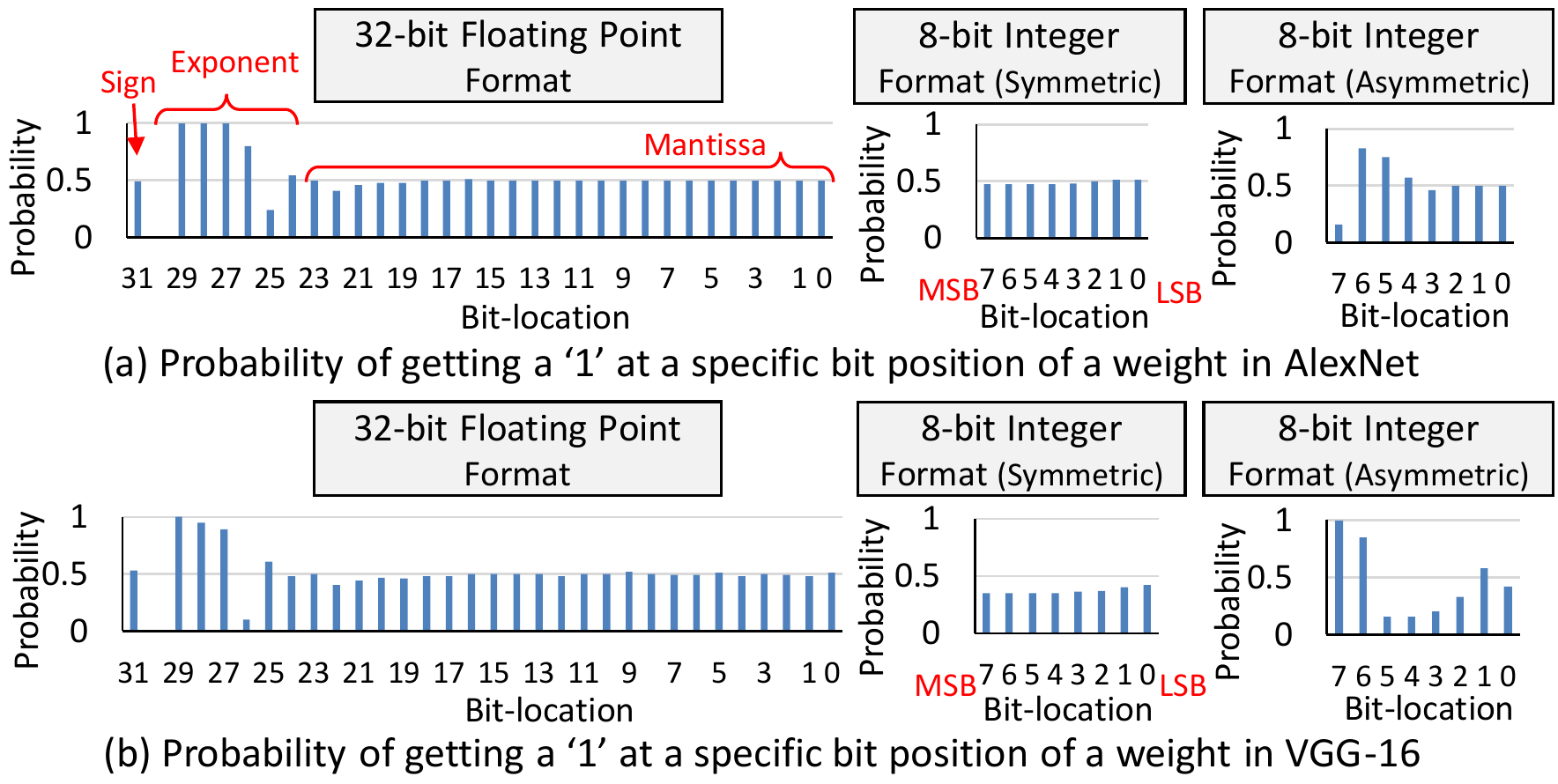}
\caption{Distribution of bits of weights of different different DNNs when represented in different data representation formats. Symmetric and asymmetric represent which post-training quantization method is used to transform the data for the corresponding distribution. \vspace{-2pt}}
\label{fig:5}
\end{figure}

For this analysis, we consider the AlexNet and the VGG-16 networks, trained on the ImageNet dataset. As different data representations for weights, we consider 32-bit floating point representation (IEEE~754 standard) and 8-bit integer format achieved using range-linear \textit{symmetric} and \textit{asymmetric} quantization techniques~\cite{lin2016fixed}. Fig.~\ref{fig:5} illustrates the 
ratio of observing a `1' to the total number of observations (which corresponds to probability of observing a `1') at each bit-location of a word for all three data representation formats for both the networks. 
\textbf{By analyzing the distributions, the following key observations are made:}

\begin{enumerate}[leftmargin=*]
    \item \textit{The probability of getting a `1' value at a particular bit-location of a randomly selected weight depends on the network, the data representation format, and the method used to transform the data to the particular data representation format.} For example, the probability of getting a `1' at a particular bit-location in symmetric 8-bit representation is almost the same across bit-locations within a network for both the considered DNNs, however, it varies across networks. Similarly, the probability of getting a `1' at the  lower bit-locations in 32-bit floating-point representation is around 0.5, however, the distribution of bits at higher bit-locations varies across bit-locations as well as across DNNs. 
    \item \textit{Representation of weights using a specific format cannot guarantee a distribution that offers 0.5 probability at each bit-location, i.e., a distribution that can potentially lead to a balanced duty-cycle.} For example, out of all the studied cases, only the distribution of the AlexNet when represented using 8-bit integer format achieved using symmetric range-linear quantization offers close to 0.5 probability for all the bit-locations. 
    \item \textit{The average probability of getting a `1' across bit-locations in a specific format is also not guaranteed to be equal to 0.5.} For example, see the distributions of 8-bit asymmetrically  quantized DNNs. Therefore, barrel shifter-based balancing techniques would not produce desirable results in such cases. 
\end{enumerate}

\subsection{A Probabilistic Model-based Analysis for Aging of 6T-SRAM \\On-chip Weight Memory of a DNN Accelerator}

In the following, we develop a probabilistic model to analyze the effectiveness of different aging mitigation techniques. 

\subsubsection{Probabilistic Model}
Assume the on-chip memory of a given DNN accelerator is composed of $I \times J$ cells. 
For mapping the weights of a DNN onto the memory, we assume: (a) the same dataflow as presented in Fig.~\ref{fig:Dataflow}; (b) each block of weights is kept in the on-chip memory for equal amount of time, and it is fetched only once during a single inference (similar to the dataflow for the DNN accelerator proposed in~\cite{delmas2018bit}); (c) each block of data mapped onto the on-chip memory fits perfectly to it. 
Based on the aforementioned conditions and the given DNN size, we can divide the DNN into $K$ blocks that translates to $K$ number of data mappings onto the on-chip weight memory. 
Now, if the same DNN is used repeatedly for inferencing with the same dataflow, a single on-chip memory cell is mapped with only $K$ different bits. 
If the probability of getting a `1' for all the bits is given by $\rho$, the probability of getting a duty-cycle less than and equal to $b/K$, or greater than and equal to $1-b/K$, can be computed using the following equation, except when $b/K = 0.5$, where the probability is 1.  

\begin{equation}\footnotesize
    P_{b/K} = \sum_{i=0}^{b} \binom{K}{i} \rho^i \times (1-\rho)^{K-i} + \sum_{i=K-b}^K \binom{K}{i} \rho^i \times (1-\rho)^{K-i}
    \label{eq:duty-cycle}
\end{equation}

Here, $b$ is an arbitrary variable with the range from $0$ to $\floor{K/2}$. 
Note that we combine (i) the cases in which duty-cycle is less than and equal to $b/K$ and (ii) the cases in which duty-cycle is greater than and equal to $1-b/K$, because in a symmetric 6T-SRAM cell both the cases cause the same level of stress in one of the two PMOS transistors. 
Assuming the above computed probability to be the same for all the cells of the on-chip memory, the probability of at least $n$ number of cells (out of $I \times J$) experiencing duty-cycle less than and equal to $b/K$, or greater than and equal to $1-b/K$ can be computed using the following equation.

\begin{equation}\footnotesize
    P_{n} = \sum_{i=n}^{I \times J} \binom{I \times J}{i} P_b^i \times (1-P_b)^{I \times J-i}
    \label{eq:10}
\end{equation}

\subsubsection{An Example Case-Study}
\begin{figure}[t]
\centering
\includegraphics[width=1\linewidth]{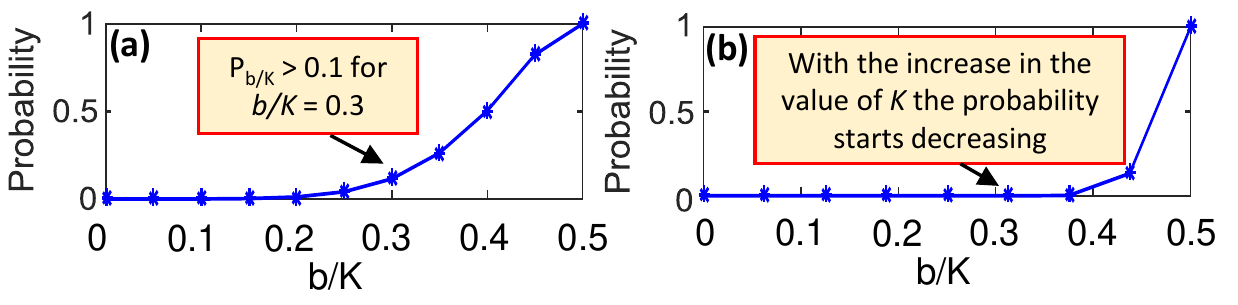}
\caption{Probability of occurrence of $b/K \geq $~duty-cycle~$\geq 1-b/K$ when (a) $K=20$, and (b) $K=160$ \vspace{-4pt}}
\label{fig:stats1}
\end{figure}

Let us consider a scenario where $K=20$ and $\rho = 0.5$ (i.e., the best-case with balanced bit distribution), and $I \times J = 8192$. Fig.~\ref{fig:stats1}a shows the probability for each possible value of $b$ computed using Eq.~\ref{eq:duty-cycle}. Note, even for $b/K = 0.3$, the probability is over 0.1, i.e., more than 10\% of the cells are expected to experience a duty-cycle of less than 0.3, or greater than 0.7.  

Now, if we employ a given aging mitigation technique that offers upto 7 shifts to increase the number of different bits that are mapped to a single cell, we can theoretically increase the value of $K$ to 160, assuming the bits to be independent from each other and the ideal shifting policy. Putting $K = 160$ in the above mentioned example, Fig.~\ref{fig:stats1}b shows the probabilities for different $b/K$ values. 
As can be seen from Fig.~\ref{fig:stats1}b, the probabilities at lower $b/K$ values have dropped significantly. \textit{The above analysis implies that by significantly increasing $K$ and having $\rho=0.5$, we can achieve close to ideal duty-cycle for all the cells. }

Now, instead of a barrel shifter, if we employ an inversion-based duty-cycle balancing technique where every other write to the same location is inverted, for the given scenario, the value of $K$ remains the same, as it is even. Moreover, as $\rho$ is defined to be 0.5, the inversion-based policy has no impact on $\rho$ either. 
Therefore, we get the same probabilities as presented in Fig.~\ref{fig:stats1}a. However, note that \textit{the inversion-based policy is mainly useful for achieving $\rho=0.5$ in cases where the distribution of bits is biased either towards `0' or `1'}. 

\subsection{Challenges in Designing an Efficient Aging Mitigation System}
\label{sec:challenges}

Based on the above analysis, we outline the following key challenges in designing a \textit{generic} aging mitigating system. 
\begin{enumerate}[leftmargin=*]
    \item The probability of occurrence of non-ideal duty-cycle is considerable even with the state-of-the-art \textit{fixed} aging mitigation techniques. Therefore, a more robust method has to be designed by exploiting the fact that NBTI-aging is more dependent on the average duty-cycle over the lifetime of the device~\cite{abella2007penelope}. 
    \item The distribution of bits and the duty-cycle is significantly affected by the datatype used for representing the weights. \textit{Therefore, the mitigation technique should be generic and independent of the datatype used so that it is beneficial for various DNN accelerators.} 
\end{enumerate}
Moreover, in practical scenarios, each layer of a DNN can have a different size. Therefore, each layer can take different amount of time for processing that can vary significantly across layers. Also, different DNNs can have different number of layers. Therefore, a method that keeps track of all these factors at a fine granularity can help in significantly reducing the aging rates. However, such methods are super costly. \textit{This makes it very challenging to develop a generic method that offers effective aging mitigation at reasonable overheads.} 

\section{A Micro-architecture for Mitigating Aging of the On-Chip Weight Memory of DNN Accelerators}
\label{sec:novel_micro_architecture}

\begin{figure}[t]
\centering
\includegraphics[width=0.82\linewidth]{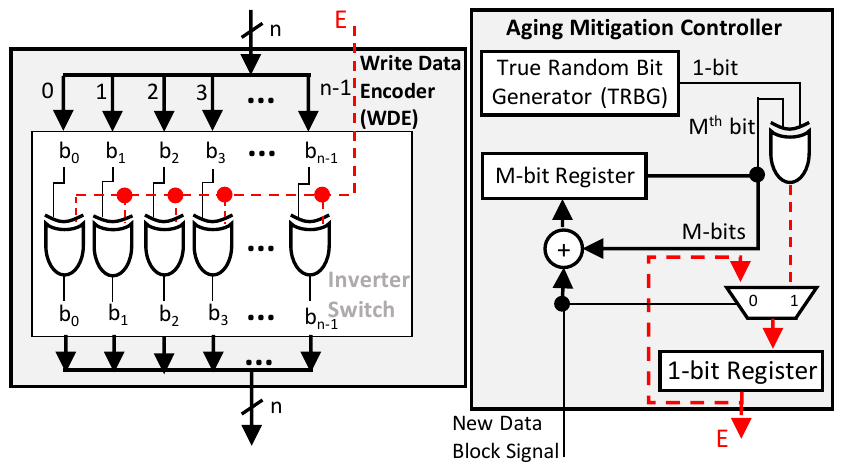}
\caption{Proposed micro-architecture for effective aging mitigation of 6T-SRAM weight memory of DNN accelerators.}
\label{fig:7}
\end{figure}

To address the above challenges, we propose a \textit{Write Data Encoder} (WDE) for encoding the weights before writing them to the on-chip weight memory, and a \textit{Read Data Decoder} (RDD) which performs the inverse function of the WDE while reading the data from the on-chip memory and before passing it to the processing array. The integration of the proposed modules in the DNN accelerator is shown in Fig.~\ref{fig:5a}a. Moreover, we propose an aging mitigation controller which generates the control signals (metadata) for the write (and read) transducer. 
The proposed micro-architectures of the WDE and the aging mitigation controller is shown in Fig.~\ref{fig:7}. 

\textbf{Write Data Encoder (WDE):} It leverages the inversion logic  
that besides its low-overhead\footnote{low overhead compared to other techniques such as shifting, which requires costly barrel shifters (as shown later in Section~\ref{Sec:Results})}, also enables to perfectly balance out the distribution of bits in the cells of the memory when the distribution is originally biased towards either `0' or `1', as highlighted in Section~\ref{sec:analysis}. 
The inversion logic in the proposed micro-architecture is implemented using XOR gates as they allow the aging mitigation controller to enable or disable it using just a \textit{1-bit enable} ($E$) signal. 
\textit{Another key advantage of this design is that the micro-architecture of the RDD is the same as WDE}, where the same $E$ signal (metadata) that is used to encode the weights is used (at a later point in time) for decoding them before passing them to the processing array. 
Moreover, the proposed WDE and RDD modules are highly \textit{scalable}, as increasing the width of the modules require only a linear increase in the number of XOR gates. Therefore, the widths of these modules can be defined directly based on the DNN accelerator configuration without affecting the energy-efficiency of the system.

\textbf{Aging Mitigation Controller:} The controller is the core part of the proposed micro-architecture, as it is responsible for generating the enable signal ($E$) that enables/disables the inversion logic in WDE. 
The design is based on the observations made in Section~\ref{sec:analysis} that the higher the number of different bits to be written on an SRAM cell during its lifetime (i.e., $K$ in Eq.~\ref{eq:duty-cycle}) that are generated from a uniform distribution the lower the chances of observing a deviation in its duty-cycle from 0.5 (see Figs.~\ref{fig:stats1}), i.e., the ideal point shown in Fig.~\ref{fig:1}b.
Therefore, to increase the number of different bits to be written on an SRAM cell, we employ a True Random Bit Generator (TRBG) to generate the enable signal and decide whether the upcoming data should be written \textit{with} or \textit{without} inversion in the memory cell. TRBG adds the sense of randomness in the bits to be written in the memory and thereby leads to larger $K$ value and lower aging. 

Note in practical scenarios, the output of TRBGs can be biased towards either `0' or `1', which can eventually affect the duty-cycle. Therefore, to mitigate this, we periodically invert the output of the TRBG after a defined number of iterations with the help of an $M$-bit register before using it as the enable signal, which balances the bias.

\begin{figure*}[t]
\centering
\includegraphics[width=1\linewidth]{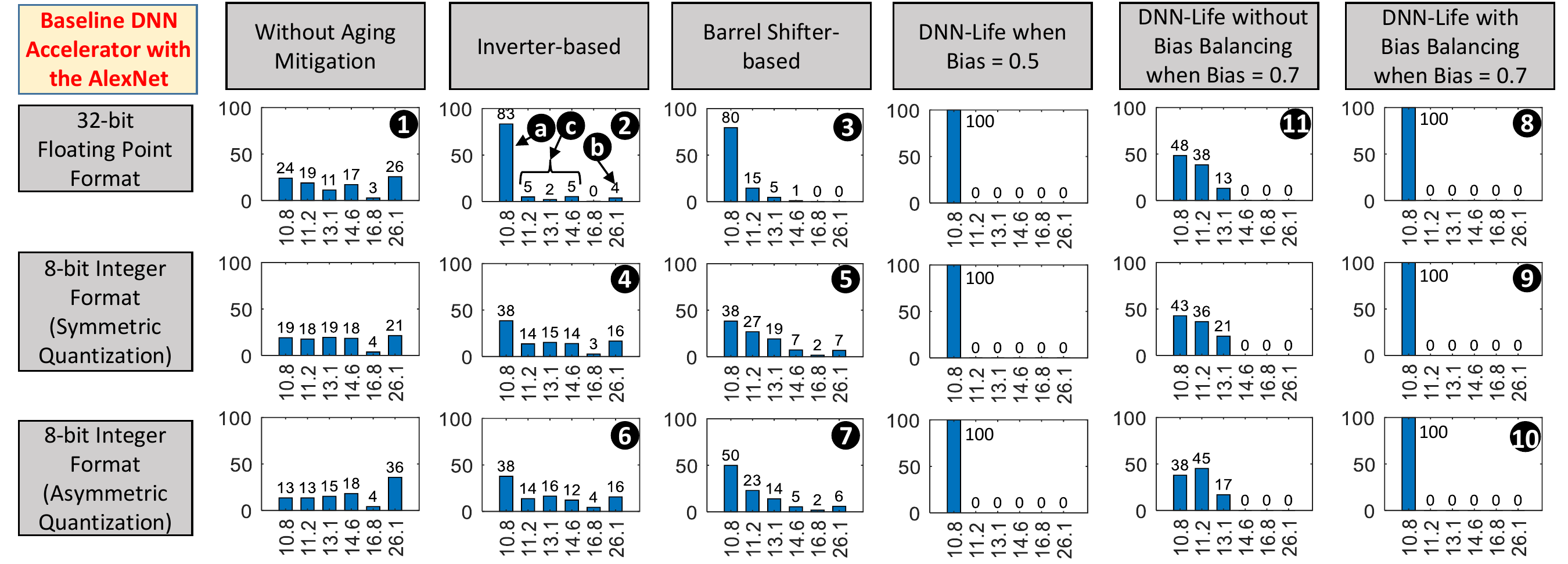}
\caption{SNM degradation of 6T-SRAM on-chip weight memory cells of the baseline DNN accelerator when used for performing inferences only using the AlexNet network. Each bar graph shows the percentage of the number of cells (Y-axis) experiencing different level of SNM degradation (X-axis).\vspace{-15pt}}
\label{fig:19}
\end{figure*}

\begin{figure}[t]
\centering
\includegraphics[width=0.9\linewidth]{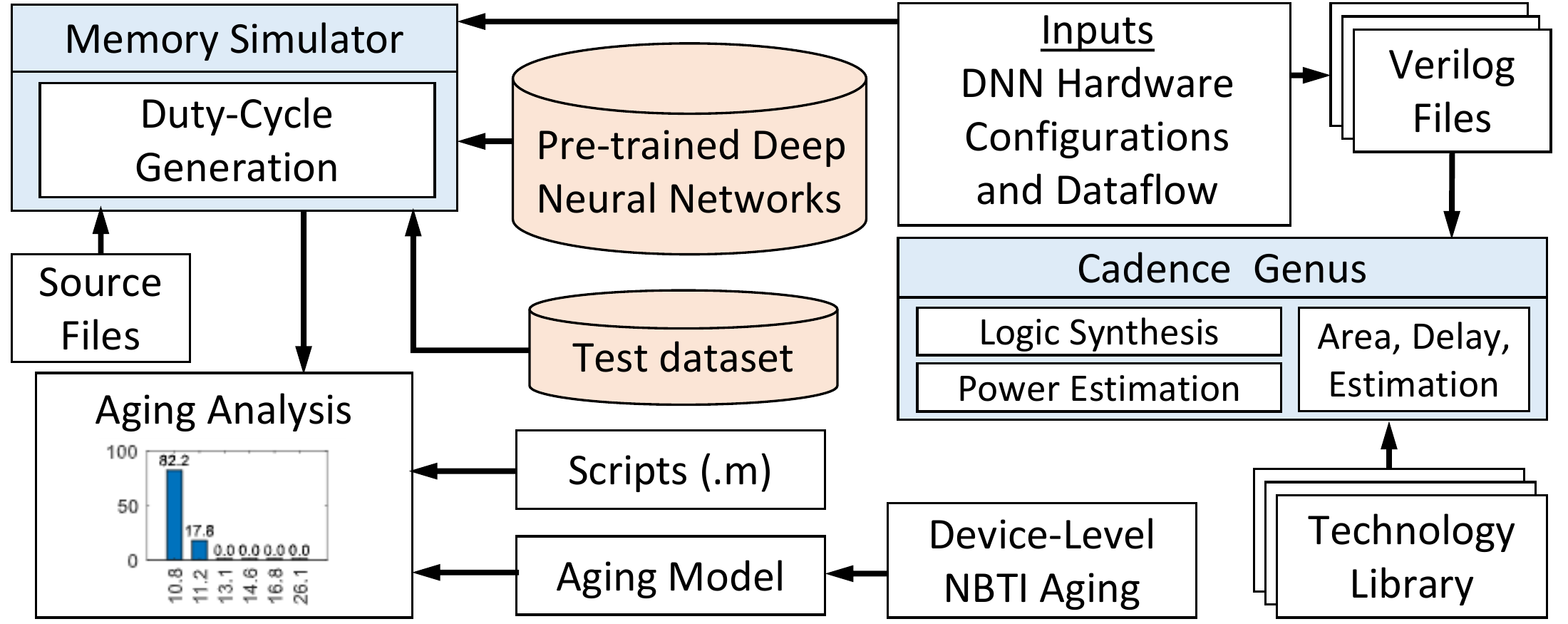}
\caption{Overall experimental setup used for evaluation.}
\label{fig:ES}
\end{figure}

\section{Results and Discussion}
\label{Sec:Results}
\subsection{Experimental Setup}
Fig.~\ref{fig:ES} illustrates the overall experimental setup used for evaluation. 
The setup consists of hardware synthesis for estimating the power, area and delay characteristics of the proposed modules, and simulations for aging estimation of the 6T-SRAM on-chip weight memory of different DNN hardware accelerators. For hardware synthesis, we implemented different aging balancing circuits and our DNN-Life architecture in Verilog. The circuits are synthesized for the TSMC 65nm technology using Cadence Genus. 

For aging estimation, we use Static Noise Margin (SNM) to quantify the NBTI-aging of 6T-SRAM cells, similar to~\cite{Shafique:2015:EEA:2744769.2744834}\cite{shafique2016content}. The SNM defines the tolerance to noise that directly affects the read stability of a cell~\cite{agarwal2006statistical}, i.e., if the SNM of a cell is low, the cell is highly susceptible to read failures. As per~\cite{kothawade2011analysis}\cite{Shafique:2015:EEA:2744769.2744834}\cite{shafique2016content}, SNM mainly depends on the duty-cycle over the entire lifetime of the cell, and the least SNM degradation is achieved at 50\% duty-cycle. To obtain SNM results, we employ a similar device aging model as used in state-of-the-art studies like~\cite{Shafique:2015:EEA:2744769.2744834}\cite{shafique2016content}. 
However, due to its duty-cycle optimization focus, our proposed technique is orthogonal to the given device aging models, and other device-level models can easily be integrated in our framework. Based on the models, the SNM degradation of a 6T-SRAM cell can be computed using the duty-cycle. From the analysis, the best SNM degradation for 6T-SRAM cell after 7 years is 10.82\% (at 50\% duty-cycle), and the worst is 26.12\% (at 0\% and 100\% duty-cycle). 

For large-scale simulations, we integrated the output of these models into a memory simulator of the baseline DNN hardware (described in Section~\ref{sec:baseline_hardware_architecture}). The simulator takes the DNN hardware configuration, dataflow, pre-trained DNN architecture and test samples as inputs. We also built a memory simulator for a TPU-like hardware architecture~\cite{jouppi2017datacenter} to validate the proposed aging-mitigation technique across DNN hardware accelerators. 
The hardware configurations used for the evaluation are presented in Table~\ref{tab:Hardware_Configs}. 
The DNNs used are the AlexNet and the VGG-16 with the ImageNet dataset and a custom network with MNIST dataset. The custom network is composed of two CONV layers and two FC layers, i.e., CONV(16,1,5,5), CONV(50,16,5,5), FC(256,800) and FC(10,256).
For each setting the duty-cycles are estimated based on the values observed in 100 inferences. The bias balancing register is defined to be a 4-bit register (i.e., M=4), for all the corresponding cases. 

\begin{table}[hb]
\caption{Hardware configurations and settings used in evaluation}
\label{tab:Hardware_Configs}
\resizebox{1\linewidth}{!}{
\begin{tabular}{l|c|c|}
\cline{2-3}
                                                                                        & \begin{tabular}[c]{@{}c@{}}Baseline Accelerator (Section~\ref{sec:baseline_hardware_architecture}) \end{tabular}                                                                        & TPU-like NPU~\cite{jouppi2017datacenter}                                                                                    \\ \hline
\multicolumn{1}{|l|}{\begin{tabular}[c]{@{}l@{}}Weight \\ memory size\end{tabular}}     & \begin{tabular}[c]{@{}c@{}}512KB\end{tabular}                         & 256KB                                                                                  \\ \hline
\multicolumn{1}{|l|}{\begin{tabular}[c]{@{}l@{}}Activation \\ memory size\end{tabular}} & 4MB                                                                                         & 24MB                                                                                   \\ \hline
\multicolumn{1}{|l|}{\begin{tabular}[c]{@{}l@{}}PE array size\end{tabular}}  & \begin{tabular}[c]{@{}c@{}}8 PEs (1 PE = 8 Multipliers)\end{tabular} & \begin{tabular}[c]{@{}c@{}}256 x 256 PEs (1 PE = 1 MAC)\end{tabular} \\ \hline
\multicolumn{1}{|l|}{Networks}                                                          & AlexNet                                                                          & \begin{tabular}[c]{@{}c@{}}AlexNet, VGG-16 and Custom\end{tabular}                 \\ \hline
\end{tabular}}
\end{table}

\subsection{Aging Estimation Results and Comparisons}
In this subsection, we analyze the impact of using different aging mitigation policies on the SNM degradation of the 6T-SRAM on-chip weight memory cells after 7 years. 
We mainly considered four different policies: (1) No aging mitigation, (2) Inversion-based, (3) Barrel shifter-based, and (4) DNN-Life. For the proposed DNN-Life, we consider three different cases: (i) TRBG is not biased and it generates 0s and 1s with equal probability (referred in the results as \textit{Bias=0.5}); (ii) TRBG is biased and it generates 1s with 0.7 probability, and the aging controller does not have a bias balancing register (referred in the results as \textit{without bias balancing with Bias=0.7}); and (iii) TRBG is biased and it generates 1s with 0.7 probability and the aging controller has a 4-bit bias balancing register (referred in the results as \textit{with bias balancing with Bias=0.7}).

Moreover, we performed experiments considering three different data representation formats for weights: (1) 32-bit floating point format; (2) 8-bit integer format when weights are quantized using symmetric quantization method; and (3) 8-bit integer format when weights are quantized using asymmetric quantization method. 

Fig.~\ref{fig:19} shows the distributions of SNM degradation in the memory cells obtained using different aging mitigation policies and a pre-trained AlexNet model. The Y-axis of each bar graph shows the percentage of the number of cells and the X-axis of each shows SNM degradation levels. 
Note that, for these experiments, we assumed the baseline DNN accelerator configuration presented in Table~\ref{tab:Hardware_Configs} and the dataflow shown in Fig.~\ref{fig:Dataflow} with $f=8$. 
Also, we assumed that only a single DNN (i.e., the AlexNet) is used for data inference throughout the lifetime of the device. 
As can be seen in the figure, the inversion-based and barrel shifter-based aging balancing reduce the SNM degradation of the SRAM cells, however, they do not offer minimum SNM degradation (see \circled{2} and \circled{3} in comparison with \circled{1} in Fig.~\ref{fig:19}). This behavior is observed to be consistent across all the data representation formats (see \circled{2} till \circled{7} in comparison with their respective \textit{without aging mitigation} graphs in Fig.~\ref{fig:19}). Specifically, the inversion-based aging balancing offers sub-optimal aging mitigation in case of the 32-bit floating point format (see \circled{2} in~Fig.~\ref{fig:19}), where most of the cells experience around 10.8\% SNM degradation (see \circled{a} in Fig.~\ref{fig:19}). However, this is not the ideal scenario as there are 4\% cells that experience highest level of SNM degradation (see \circled{b} in Fig.~\ref{fig:19}) and a few that experience moderate level of SNM degradation (see \circled{c} in Fig.~\ref{fig:19}). 
Now, if we analyze the results of the proposed \textit{DNN-Life with bias balancing}, it offers maximum aging-mitigation (i.e., all the cells experience around 10.8\% SNM degradation) in all the cases (see \circled{8}, \circled{9} and \circled{10} in Fig.~\ref{fig:19}).

\textbf{Impact of biased TRBG on aging balancing of 6T-SRAM on-chip weight memory:} Fig.~\ref{fig:19} also illustrates the impact of using proposed design without bias correction when the duty-cycle of TRBG is 0.7. As can be seen in the figure, for all the data representation formats, having biased TRBG and no bias correction leads to less reduction in SNM degradation of the 6T-SRAM cells (e.g., see \circled{11} in comparison with \circled{8} in~Fig.~\ref{fig:19}). This behavior is consistent across all the data representation formats. 

\begin{figure}[t]
\centering
\includegraphics[width=1\linewidth]{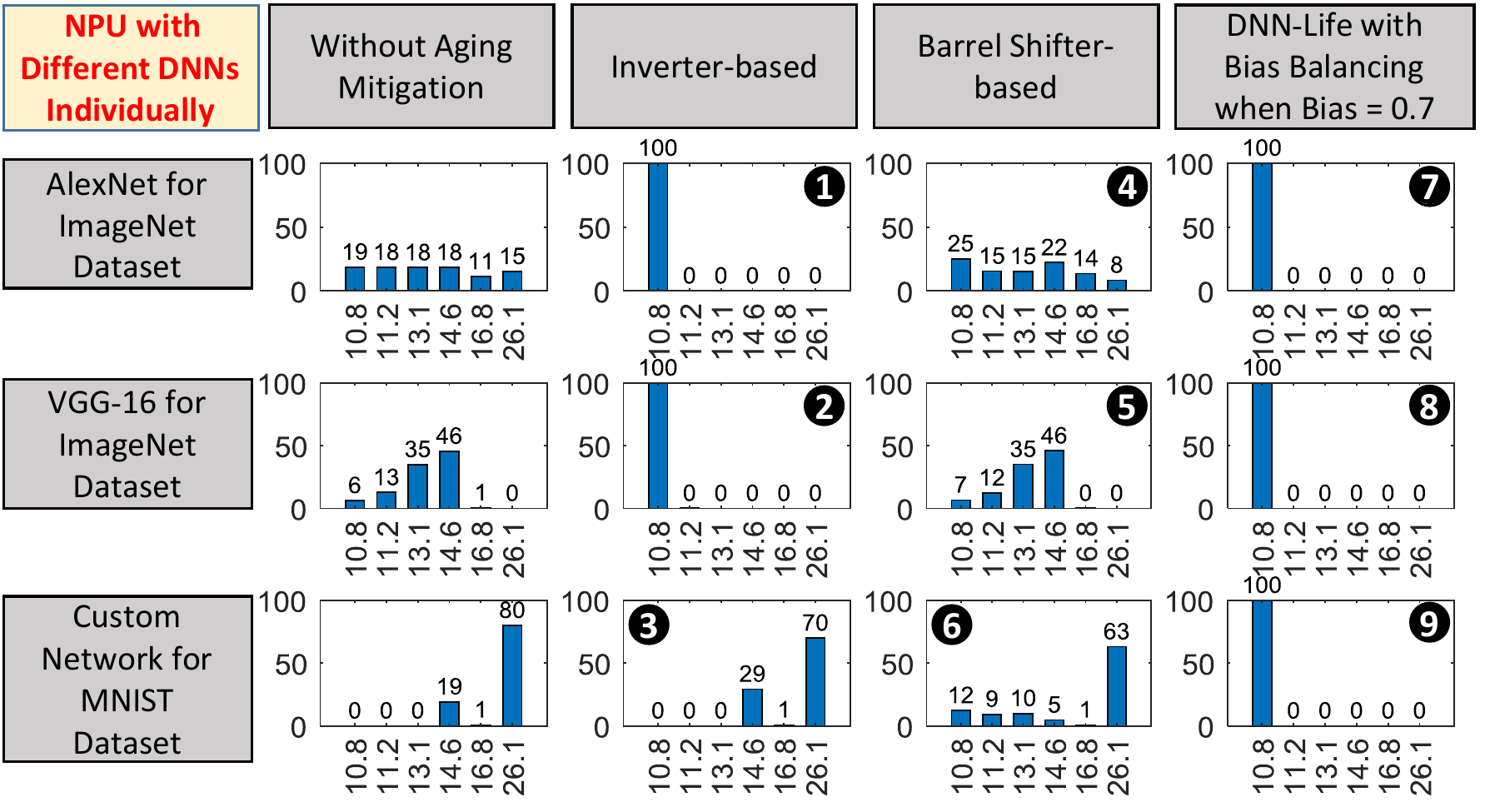}
\caption{SNM degradation of 6T-SRAM on-chip weight memory cells of a TPU-like NPU when used for performing inferences using the AlexNet, the VGG-16 and the custom DNN, individually. The networks are quantized to 8-bit format using symmetric range-linear quantization method. \vspace{-2pt}}
\label{fig:30}
\end{figure}

\textbf{Impact across different hardware accelerators:} Fig.~\ref{fig:30} shows the impact of using the proposed aging-mitigation technique for a TPU-like~\cite{jouppi2017datacenter} Neural Processing Unit (NPU) architecture that has an on-chip weight FIFO which is four tiles deep, where one tile is equivalent to weights for $256 \times 256$ PEs. Each PE has a single MAC unit that can perform 8-bit multiplication. For our implementation, we assumed the weight FIFO to be a circular buffer-based design. 
We performed analysis using the three different networks mentioned earlier. All the DNNs are quantized to 8-bits using post-training symmetric quantization. Considering the dataflow of the NPU, the parameter $f$ was set to 256. As can be seen in Fig.~\ref{fig:30}, the inversion-based aging mitigation policy offers optimal results for the AlexNet and the VGG-16 networks (see \circled{1} and \circled{2} in Fig.~\ref{fig:30}). However, when used for the custom DNN, almost all the memory cells experience significant SNM degradation (see \circled{3} in Fig.~\ref{fig:30}). The barrel shifter-based approach also offer sub-optimal results (see \circled{4} till \circled{6} in Fig.~\ref{fig:30}). However, the proposed DNN-Life with bias balancing offers maximum aging mitigation (see \circled{7} till \circled{9} in Fig.~\ref{fig:30}). This shows that DNN-Life can be used for a wide range of DNN accelerators. 

\subsection{Area and Power Results}
The area, power and delay characteristics of three different WDEs composed of different aging balancing units are shown in Table~\ref{tab:11}. 
All three WDEs are designed for 64 bit-width. 
The barrel shifter-based WDE consumes the most amount of area and power. The proposed design consumes slightly more power and area as compared to the inversion-based WDE. However, as shown in the previous subsection, it offers best aging-mitigation in all the possible scenarios regardless of the size of the given DNN, the data representation format and the on-chip weight memory size. Note that, at hardware level, we realized TRBG using a 5-stage ring oscillator. 

\begin{table}[ht]
\caption{Hardware results of different Write Data Encoders (WDEs)}
\label{tab:11}
\resizebox{1\linewidth}{!}{
\begin{tabular}{l|l|l|l|}
\cline{2-4}
                                               & Delay {[}ps{]} & Power {[}nW{]} & \begin{tabular}[c]{@{}l@{}}Area {[}cell area{]}\end{tabular} \\ \hline
\multicolumn{1}{|l|}{Barrel Shifter based WDE} & 977.7          & 345190         & 9035                                                            \\ \hline
\multicolumn{1}{|l|}{Inversion based WDE}      & 811.6          & 10716          & 195                                                             \\ \hline
\multicolumn{1}{|l|}{\begin{tabular}[c]{@{}l@{}}Proposed WDE with Aging\\Mitigation Controller\end{tabular}}   & 581.8          & 13747          & 295                                                             \\ \hline
\end{tabular}}
\end{table}

\section{Conclusion}
\label{sec:conclusion}
In this paper, we proposed DNN-Life, an aging-mitigation framework that employs read and write transducers to reduce NBTI-induced aging of 6T-SRAM on-chip weight memory in DNN hardware accelerators. We analyzed different DNN data representation formats at the software-level and their potential for balancing the duty-cycle in SRAM cells. Based on the analysis, we proposed a micro-architecture that makes use of a True Random Bit Generator (TRBG) to ensure optimal duty-cycle at runtime, thereby balancing the aging of complimentary parts in 6T-SRAM cells of the weight memory. As a result, our DNN-Life enables efficient aging mitigation of weight memory of a given DNN hardware with minimal energy overhead. 

\section*{Acknowledgment}
This work is partially supported by Intel Corporation through Gift funding for the project "Cost-Effective Dependability for Deep Neural Networks and Spiking Neural Networks."

\def\bibfont{\footnotesize}
\bibliographystyle{IEEEtran}
\bibliography{biblio}

\begin{thebibliography}{10}
\providecommand{\url}[1]{#1}
\csname url@samestyle\endcsname
\providecommand{\newblock}{\relax}
\providecommand{\bibinfo}[2]{#2}
\providecommand{\BIBentrySTDinterwordspacing}{\spaceskip=0pt\relax}
\providecommand{\BIBentryALTinterwordstretchfactor}{4}
\providecommand{\BIBentryALTinterwordspacing}{\spaceskip=\fontdimen2\font plus
\BIBentryALTinterwordstretchfactor\fontdimen3\font minus
  \fontdimen4\font\relax}
\providecommand{\BIBforeignlanguage}[2]{{%
\expandafter\ifx\csname l@#1\endcsname\relax
\typeout{** WARNING: IEEEtran.bst: No hyphenation pattern has been}%
\typeout{** loaded for the language `#1'. Using the pattern for}%
\typeout{** the default language instead.}%
\else
\language=\csname l@#1\endcsname
\fi
#2}}
\providecommand{\BIBdecl}{\relax}
\BIBdecl

\bibitem{Sze_Survey}
V.~{Sze et al.}, ``Efficient processing of deep neural networks: A tutorial and
  survey,'' \emph{Proceedings of IEEE}, vol. 105, no.~12, pp. 2295--2329, 2017.

\bibitem{capra2020hardware}
M.~{Capra et al.}, ``Hardware and software optimizations for accelerating deep
  neural networks: Survey of current trends, challenges, and the road ahead,''
  \emph{IEEE Access}, 2020.

\bibitem{chen2014dadiannao}
Y.~Chen~et al., ``Dadiannao: A machine-learning supercomputer,'' in
  \emph{IEEE/ACM \textbf{MICRO} Symposium}, 2014, pp. 609--622.

\bibitem{jouppi2017datacenter}
N.~P. Jouppi~et al., ``In-datacenter performance analysis of a tensor
  processing unit,'' in \emph{ACM/IEEE \textbf{ISCA}}, 2017, pp. 1--12.

\bibitem{HOTCHIP}
\BIBentryALTinterwordspacing
P.~McLellan. (2019) Hot chips: The biggest chip in the world. Accessed:
  2019-09-10. [Online]. Available:
  \url{https://community.cadence.com/cadence\_blogs\_8/b/breakfast-bytes/posts/the-biggest-chip-in-the-world}
\BIBentrySTDinterwordspacing

\bibitem{henkel2013reliable}
J.~Henkel~et al., ``Reliable on-chip systems in the nano-era: Lessons learnt
  and future trends,'' in \emph{ACM/ESDA/IEEE \textbf{DAC}}, 2013, p.~99.

\bibitem{shafique2020robust}
M.~{Shafique et al.}, ``Robust machine learning systems: Challenges, current
  trends, perspectives, and the road ahead,'' \emph{IEEE Design \& Test},
  vol.~37, no.~2, pp. 30--57, 2020.

\bibitem{henkel2013thermal}
J.~Henkel~et al., ``Thermal management for dependable on-chip systems,'' in
  \emph{IEEE \textbf{ASP-DAC}}, 2013, pp. 113--118.

\bibitem{8474192}
M.~A. {Hanif et al.}, ``Robust machine learning systems: Reliability and
  security for deep neural networks,'' in \emph{IEEE IOLTS}, 2018, pp.
  257--260.

\bibitem{kim2018matic}
S.~Kim~et al., ``Matic: Learning around errors for efficient low-voltage neural
  network accelerators,'' in \emph{IEEE \textbf{DATE}}, 2018, pp. 1--6.

\bibitem{kang2008nbti}
K.~Kang~et al., ``Nbti induced performance degradation in logic and memory
  circuits: How effectively can we approach a reliability solution?'' in
  \emph{IEEE \textbf{ASP-DAC}}, 2008, pp. 726--731.

\bibitem{gnad2015hayat}
D.~{Gnad et al.}, ``Hayat: Harnessing dark silicon and variability for aging
  deceleration and balancing,'' in \emph{2015 52nd ACM/EDAC/IEEE DAC}, 2015.

\bibitem{shin2008proactive}
J.~Shin~et al., ``A proactive wearout recovery approach for exploiting
  microarchitectural redundancy to extend cache sram lifetime,'' in
  \emph{ACM/IEEE SIGARCH Computer Arch. News}, vol.~36, no.~3, 2008, pp.
  353--362.

\bibitem{abella2007penelope}
J.~Abella~et al., ``Penelope: The nbti-aware processor,'' in \emph{IEEE/ACM
  \textbf{MICRO} Symposium}.\hskip 1em plus 0.5em minus 0.4em\relax IEEE
  Computer Society, 2007, pp. 85--96.

\bibitem{kothawade2011analysis}
S.~Kothawade~et al., ``Analysis and mitigation of nbti aging in register file:
  An end-to-end approach,'' in \emph{IEEE \textbf{ISQED}}, 2011, pp. 1--7.

\bibitem{ricketts2010investigating}
A.~Ricketts~et al., ``Investigating the impact of nbti on different power
  saving cache strategies,'' in \emph{IEEE \textbf{DATE}}, 2010, pp. 592--597.

\bibitem{siddiqua2011enhancing}
T.~Siddiqua~et al., ``Enhancing nbti recovery in sram arrays through recovery
  boosting,'' \emph{IEEE \textbf{TVLSI}}, vol.~20, no.~4, pp. 616--629, 2011.

\bibitem{6105303}
B.~{Zatt et al.}, ``A low-power memory architecture with application-aware
  power management for motion disparity estimation in multiview video coding,''
  in \emph{IEEE/ACM \textbf{ICCAD}}, 2011, pp. 40--47.

\bibitem{jin2012aging}
T.~Jin~et al., ``Aging-aware instruction cache design by duty cycle
  balancing,'' in \emph{IEEE \textbf{IVLSI}}, 2012, pp. 195--200.

\bibitem{calimera2011partitioned}
A.~Calimera~et al., ``Partitioned cache architectures for reduced nbti-induced
  aging,'' in \emph{IEEE \textbf{DATE}}, 2011, pp. 1--6.

\bibitem{Shafique:2015:EEA:2744769.2744834}
M.~Shafique~et al., ``Enaam: Energy-efficient anti-aging for on-chip video
  memories,'' in \emph{ACM/IEEE \textbf{DAC}}, 2015, pp. 101:1--101:6.

\bibitem{delmas2018bit}
A.~Delmas~et al., ``Bit-tactical: Exploiting ineffectual computations in
  convolutional neural networks: Which, why, and how,'' \emph{preprint
  arXiv:1803.03688}, 2018.

\bibitem{8342033}
J.~{Li et al.}, ``Smartshuttle: Optimizing off-chip memory accesses for deep
  learning accelerators,'' in \emph{IEEE \textbf{DATE}}, 2018, pp. 343--348.

\bibitem{lin2016fixed}
D.~Lin~et al., ``Fixed point quantization of deep convolutional networks,'' in
  \emph{\textbf{ICML}}, 2016, pp. 2849--2858.

\bibitem{shafique2016content}
M.~{Shafique et al.}, ``Content-aware low-power configurable aging mitigation
  for sram memories,'' \emph{IEEE Transactions on Computers}, vol.~65, no.~12,
  pp. 3617--3630, 2016.

\bibitem{agarwal2006statistical}
K.~Agarwal~et al., ``Statistical analysis of sram cell stability,'' in
  \emph{ACM/IEEE \textbf{DAC}}, 2006, pp. 57--62.

\end{thebibliography}

\end{document}